# Absorption and Emission Probabilities of Electrons in Electric and Magnetic Fields for FEL

I.V. Dovgan*

Department of Physics, Moscow State Pedagogical University, Moscow 119992, Russia.

We consider induced emission of ultrarelativistic electrons in strong electric (magnetic) fields that are uniform along the direction of the electron motion and are not uniform in the transverse direction. The stimulated absorption and emission probabilities are found in such system.

1. **INTRODUCTION**

The motion and radiation of relativistic electrons in stationary spatially periodic magnetic fields (undulators) has been investigated in detail in connection with the problem of free-electron lasers (FEL) [1-2]. One such method may be the use of ultrarelativistic electrons that are uniform along the direction of motion but are not uniform in a direction transverse to the stationary electric and magnetic fields. If the field potential in such fields increases in a direction perpendicular to the beam from the center to the periphery, the electrons can execute harmonic oscillations. In [3-5] transverse channeling of electrons in intense standing light wave was presented. Many aspects of amplification in FEL are presented in [6-71].

We consider in this paper stimulated emission of relativistic electrons in strong electric and magnetic fields that are uniform along the electron-motion direction and are not uniform in the transverse direction. It is assumed that the potential energy of the interaction of the electrons with the field depends quadratically on the transverse coordinates. We shall assume that a one-dimensional case is realized, when the field varies only along one of the axes in the transverse direction. A generalization of the solution to the two dimensional case does not lead to results that are fundamentally new.

**2. BASIC EQUATIONS**

The static field in which the relativistic electron move will be specified via the 4-potential $A_1 = (\Phi_1, \mathbf{A}_1)$, whose value is assumed to vary along the Z axis in accord with the quadratic law $A_1(z) = A_{01}(z^2/d^2)$, where $A_{01}$ is the maximum amplitude of the 4-potential and $-d/2 < z < d/2$.

---

*dovganirv@gmail.com

We assume the field to be uniform in the direction of motion of the electron beam (the X axis) and in the perpendicular direction (the Y axis).

Let $p_\parallel \gg p_\perp$, where $p_\parallel$ and $p_\perp$, are respectively the longitudinal and transverse components of the initial momentum of the electron. In accordance with the considered character of the particle motion, we represent the total initial energy $\varepsilon$ by the sum $\varepsilon \approx \varepsilon_\parallel + \varepsilon_\perp$, w here $\varepsilon_\parallel = (p_\parallel^2 c^2 + m_e^2 c^4)^{1/2}$ and $\varepsilon_\perp = p_\perp^2 / 2\gamma m_e$ ( $\gamma = \varepsilon / m_e c^2$ is the relativistic factor and $m_e$ is the electron mass) are the energies corresponding to motion longitudinal and transverse relative to the Z axis. We assume satisfaction of the strong inequality $\varepsilon \approx \varepsilon_\parallel \gg \varepsilon_\perp$.

The field of a traveling electromagnetic field is defined by the 4-potetial ($\hbar = c = 1$)

$$A_2(x) = \frac{1}{2} A_{02} \left[ e_2 e^{-ikx} + c.c. \right], \tag{1}$$

where $e_2 = (0, \mathbf{e}_2)$ is a unit vector of the wave-field polarization; $A_{02}$ is the amplitude of the field; $k = (\omega, \mathbf{k})$ is the 4- momentum of the field quantum. We have used in (1) the usual notation for the scalar product of 4-vectors: $kx = (kx) = \omega t - \mathbf{k} \cdot \mathbf{r}$.

As the basic equation, neglecting small spin correction, we use the Klein-Gordon equation in the fields $A_{1,2}(x)$. The dimensionless parameter $K = eA_{01} / m_e$, which characterizes the intensity of the interaction of the electron with the electric (magnetic) field can be $\geq 1$, and accordingly we take the field $A_1(x)$ into account in all orders of perturbation theory. The wave field $A_2(x)$ is assumed weak enough and we consider it in first-order perturbation theory. The Klein-Gordon equation takes in a field with potential $A_1$ the form

$$\left[ -\frac{\partial}{\partial x_\mu} \frac{\partial}{\partial x^\mu} - 2ie\left( A_1^\mu \frac{\partial}{\partial x^\mu} \right) + (eA_1)^2 - m_e^2 \right] \Psi - 0. \tag{2}$$

We seek the solution of (2) in the form [7]

$$\Psi(\mathbf{r}, t) = \frac{1}{\sqrt{2\varepsilon V}} e^{-ipx} F(z, t), \tag{3}$$

Where $p = (\varepsilon, \mathbf{p})$ is the 4-momentum of the free electron (when the field is turned off). In (3) we use the usual normalization to a single particle in the volume V.

The form of the subsequent formulas depends on the type of the field. Thus, in the case of an electrostatic field ($A_1 = 0$) Eq. (2) takes the form

$$\left[ -\frac{\partial^2}{\partial t^2} + \Delta - 2ie\Phi_1 \frac{\partial}{\partial t} + e^2 \Phi_1^2 - m_e^2 \right] \Psi - 0. \tag{4}$$

($\Phi_1$ is the scalar potential of the field); in the case of a magnetic field ($\Phi_1 = 0$) Eq. (2) is written in the form

$$\left[ -\frac{\partial^2}{\partial t^2} + \Delta - 2ie(\mathbf{A}_1 \nabla) - e^2 A_1^2 - m_e^2 \right] \Psi - 0. \tag{5}$$

($\mathbf{A}_1$ is the vector potential of the field). Since the general method of solving Eqs. (4) and (5), as well as the approximations used; are analogous in this approximation, we confine ourselves hereafter to the electrostatic field and will present without derivation the results for a stationary magnetic field.

Substitution of (3) in (4) leads to an equation of the Schrodinger type for the transverse motion of the particle:

$$i \frac{\partial \tilde{F}}{\partial t} = \left[ -\frac{1}{2\gamma m_e} \frac{\partial^2}{\partial z^2} + e\Phi_1(z) \right] \tilde{F}. \tag{6}$$

The wave function of the particle (3) breaks up into factors that correspond to its longitudinal and transverse motion

$$\Psi(\mathbf{r}, t) = \frac{1}{\sqrt{2\varepsilon S}} e^{-i(\varepsilon_\parallel t - \mathbf{p}_\parallel \mathbf{\rho})} \tilde{F}(z, t), \tag{7}$$

(in accord with the representation of the function (7), it is convenient to express the volume V in the form V = SL, where S is the normalization area in the XY plane, L is the normalization length in the direction of the Z axis, and $\rho$ is the radius vector of the particle in the XY plane)

In the derivation of (6) we left out the small terms $\frac{\partial^2 \tilde{F}}{\partial t^2}$ and $e^2\Phi_1^2 F$, this is valid if the inequalities $\varepsilon \gg \varepsilon_\perp$ and $\varepsilon \gg V_{0E}$ are satisfied ($V_{0E} = e\Phi_{01}$ is the depth of the potential well in which the transverse motion of the electrons takes place. The inequality $V_{0E}/\varepsilon \ll 1$ can be formulated in the form of the condition on the field intensity and the particle energy, namely $K/\gamma \ll 1$ ($K = e\Phi_{01}/m_e$).

The eigensolutions of the stationary Schrodinger equation

$$-\frac{1}{2\gamma m_e}\frac{\partial^2 \varphi_n}{\partial z^2} + e\Phi_{01}\frac{z^2}{d^2}\varphi_n = \varepsilon_n \varphi_n$$

form an orthonormalized basis of oscillator functions with energy eigenvalues $\varepsilon_n = (n+1/2)\Omega_E$, where

$$\Omega_E = \frac{1}{d}\left(\frac{2e\Phi_{01}}{\gamma m_e}\right)^{1/2} = \frac{1}{d}\left(\frac{2V_{0E}}{\varepsilon}\right)^{1/2}. \tag{8}$$

To solve the nonstationary equation (6) we assume that the interaction of the electron with the field is turned on instantaneously at the instant of time t = 0 (the corresponding condition for the entry of the particle into the field will be formulated below). The basic function that describes the state of the particle at instants t > 0 is then

$$\Psi_i(\mathbf{r},t) = \frac{1}{\sqrt{2\varepsilon S}} e^{-i(\varepsilon_\parallel t - \mathbf{p}_\parallel \boldsymbol{\rho})} \sum_n a_n \varphi_n(z) e^{-i\Omega_E (n+1/2)t}, \tag{9}$$

where the expansion coefficients are given by

$$a_n = \int \tilde{F}(z,0)\varphi_n(z)dz \tag{10}$$

and can be interpreted as the probability amplitudes of the transition of the particle from free-motion states $\tilde{F}(z, t \le 0)$ into definite oscillator states $\varphi_n(z)$ when the field is suddenly turned on.

We formulate now the condition for the suddenness of turning on the interaction. This approximation takes place in the case when the characteristic time of entry of the particle in the field $\tau \sim d/v \approx d$ (v is the electron velocity) is much shorter than the period $T \sim (\varepsilon_{n+1} - \varepsilon_n)^{-1}$ of the natural oscillations of the system, i.e., when the inequality $d \ll \pi/\Omega_E$ is satisfied. It is easily seen that this criterion can be formulated in the form of a condition on the value of the parameter K and on the particle energy $(K/\gamma)^2 \ll 1$.

We make now a few remarks concerning the form of the function (9) and of the coefficients (10). Since the potential well for the transverse motion has a finite depth, the complete basis of the eigenfunctions contains, strictly speaking not only the oscillator functions $\varphi_n(z)$ but also continuous spectrum functions. It is therefore legitimate to retain the discrete states in the expansion (9) only when the energy of the transverse motion of the particle on entering the field does not exceed the height of the potential barrier: $\varepsilon_\perp \leq V_{0E}$. When this condition is satisfied the well is filled with states that are not too close in energy to the height of the barrier, the summation in (9) extends to the value n = n,, corresponding to the upper level ($n_0 = V_{0E}/\Omega_E$). The restriction on the value of $\varepsilon_\perp$ can be formulated in the form of a condition on the angle $\Theta_0 = p_\perp/\varepsilon$ of entry of the particle into the field:

$$\Theta_0 \leq (2V_{0E}/\varepsilon)^{1/2}. \tag{I1}$$

We turn now to the coefficients $a_n$, which give the populations of the levels in the well. In accord with the meaning of the expansion (9), these coefficient should satisfy the normalization condition $\sum_n |a_n|^2 = 1$. To satisfy this condition we use for the function $\tilde{F}(z, t \leq 0)$ normalization in a volume with side $L \gg 1/p_\perp$, so that in accordance with the definition (10) the coefficients are given by the formula

$$a_n = \frac{1}{\sqrt{L}} \int_{-L/2}^{L/2} e^{ip_\perp z} \varphi_n(z) dz. \tag{12}$$

The wave function of the final state of the particle is obtained from (9) and (10) by formal replacement of all the initial parameters by the final ones:

$$\Psi_i(\mathbf{r},t) = \frac{1}{\sqrt{2\varepsilon' S}} e^{-i(\varepsilon'_\| t - \mathbf{p}'_\| \boldsymbol{\rho})} \sum_m a'_m \varphi_m(z) e^{-i\Omega_E(m+1/2)t},$$

$$a'_m = \frac{1}{\sqrt{L}} \int_{-L/2}^{L/2} e^{ip_\perp z} \varphi_m(z) dz. \quad (13)$$

In the case of a magnetostatic field the Schrodinger equation that describes the transverse motion of the particle is obtained from (6) by making the obvious replacement of the potential energy $e\Phi_1(z)$ by $-eA_1(z)(\mathbf{pe}_1)/\gamma m_e$, where $\mathbf{e}_1$ is a unit vector of the magnetic-field polarization and is perpendicular to the z axis. The natural frequency of the system is then equal to

$$\Omega_M = \frac{(2eA_{01}|\mathbf{pe}_1|)^{1/2}}{d\gamma m_e} = \frac{1}{d}\left(\frac{2V_{0E}}{\varepsilon}\right)^{1/2}. \quad (14)$$

and depends parametrically not only on the magentic-field amplitude but also on the energy and direction of motion of the particle. Owing to this dependence it is necessary to formulate an additional criterion, upon satisfaction of which we obtain for the magnetic field a Schrodinger equation of the type (6): $|\mathbf{pe}_1| \gg eA_{01}$. Obviously, this condition restricts the possible directions of the longitudinal component $\mathbf{p}_\|$ of the particle momentum relative to the vector $A_{01}$.

As noted above, the field $A_2(x)$ of a traveling electromagnetic wave field is taken into account in first-order perturbation theory. The perturbation operator linear in the wave field takes for the case of an electrostatic field the form

$$\hat{V}_E = -2ie(\mathbf{A}_2 \nabla), \quad (15)$$

and for a magneto static field

$$\hat{V}_E = -2ie(\mathbf{A}_2 \nabla) - 2e^2(\mathbf{A}_1 \mathbf{A}_2). \quad (16)$$

Again, as above, we confine ourselves to the electrostatic field. The results for the magnetic field will be given in final form without proof.

The processes considered in this paper are characterized by an S-matrix element the expression for which in first. order perturbation theory in terms of the field $A_2(x)$ is of the form

$$S_{fi} = -\frac{i}{2\sqrt{\varepsilon\varepsilon'}S} \sum_{m,n} a_m^{'*} a_n \int e^{i(\varepsilon_\parallel' + \Omega m \pm \omega - \varepsilon_\parallel - \Omega n)t} e^{i(\mathbf{p}_\parallel' \pm \mathbf{k}_\parallel - \mathbf{p}_\parallel)\boldsymbol{\rho}} d\boldsymbol{\rho} dt \qquad (17)$$

(in this and following formulas we leave out the subscript $E$, which is of no in the derivation; the upper and lower signs correspond respectively to emission and absorption).

We assume an infinite region of interaction between the electrons in the field (more accurately, we assume satisfaction of the condition $\delta\varepsilon/\varepsilon > 2\pi/l\Omega$, where $\delta\varepsilon/\varepsilon$ is the relative energy scatter in the initial electron beam and l is the linear dimension of the interaction region. The parameter $l\Omega/2\pi$ determines the effective number of the electron oscillations in the transverse direction over a length l. In this case the integrations in (17) are between infinite limits, and as a result we obtain a product of three $\delta$-functions connected with the system energy and momentum conservation laws for longitudinal motion. In the upshot we obtain from (17) the following expression for the S-matrix elements that describe processes with emission ( $S_{fi}^{(e)}$ ) and absorption ( $S_{fi}^{(a)}$ ) of a wave quantum with energy $\omega$:

$$\begin{aligned}
S_{fi}^{e(a)} &= -\frac{i}{2\sqrt{\varepsilon\varepsilon'}S}(2\pi)^3 \delta^{(2)}(\mathbf{p}_\parallel' \pm \mathbf{k}_\parallel - \mathbf{p}_\parallel) \\
&\times \sum_{m,n} a_m^{'*} a_n \delta[\varepsilon_\parallel' + \Omega m \pm \omega - \varepsilon_\parallel - \Omega n]\{e(\mathbf{A}_{02}\mathbf{p}_\parallel)I_{m,n}^{(1)} - ie(A_{02})_z I_{m,n}^{(2)}\}; \\
I_{m,n}^{(1)} &= \int \varphi_m^*(z) e^{\mp ik_\perp z} \varphi_n(z) dz, \\
I_{m,n}^{(2)} &= \int \varphi_m^*(z) e^{\mp ik_\perp z} \frac{d\varphi_n(z)}{dz} dz.
\end{aligned} \qquad (18)$$

From the conservation laws contained in the $\delta$ functions of (18) follows a relation for the frequency $\omega$ of the wave for which amplification is possible

$$\omega_s = \frac{s\Omega}{1-(p_\parallel/\varepsilon_\parallel)\cos\theta} \approx \frac{2\gamma^2 s\Omega}{1+\gamma^2\theta^2}, \qquad (19)$$

where $s = |n-m|$; $\theta$ is a small angle between the vectors $\mathbf{k}$ and $\mathbf{p}_\parallel$.

When transforming to probabilities of the processes, the singularity that remains after integration with respect to dp' in (18), as is customary in problems of induced emission in a given field, is eliminated by subsequent integration that takes into account, for example, the finite character of the interaction region, the anharmonicity of the field, etc. in our formulation, the total probabilities of the processes should be averaged over the initial energy distribution of the electrons in the beam, given by the distribution function $f(\varepsilon)$. We assume that this function is normalized to unity by the condition $\int f(\varepsilon)d\varepsilon = 1$ and that the width of the function $\delta\varepsilon \ll \varepsilon$.

Taking all the foregoing into account, the probabilities per unit time of the processes with emission ($dw_e$) and absorption ($dw_a$) of a quantum $\omega$ are given by the expressions

$$dw_{e,a} = \frac{\pi}{2} \sum_{m,n} |a_n|^2 \frac{\delta[\varepsilon_{\parallel}^{'} \pm \omega - \varepsilon_{\parallel} - \Omega(n-m)]}{\varepsilon\varepsilon_{e,a}^{'}} \quad (20)$$
$$\times |e(\mathbf{A}_{02}\mathbf{p}_{\parallel})I_{m,n}^{(1)} - ie(A_{02})_z I_{m,n}^{(2)}|^2 f(\varepsilon)d\varepsilon.$$

Equation (20) was obtained as a result of integration over the phase space of the scattered particle, using the sum rule for coefficients a, and the condition that the functions $\varphi_n(z)$ are orthonormalized. Using the known expressions for the integrals $I_{m,n}^{(1)}$ and $I_{m,n}^{(2)}$ in terms of Laguerre polynomials [8] obtain from (20)

$$dw_e = \frac{\pi}{2} \frac{\delta(\varepsilon_{\parallel e}^{'} \pm \omega_e - \varepsilon_{\parallel} - s\Omega)}{\varepsilon\varepsilon_{e,a}^{'}} \sum_{n=s} |a_n|^2 \frac{(n-s)!}{n!} e^{-\alpha} \left\{ e(\mathbf{A}_{02}\mathbf{p}_{\parallel})\alpha^{3/2}L_n^3(\alpha) \right.$$
$$\left. + \frac{1}{\sqrt{2}} \frac{e(A_{02})_z}{z_0} [n\alpha^{(s-1)/2}L_{n-1}^{s-1}(\alpha) + \alpha^{(s+1)/2}L_{n+1}^{s+1}(\alpha)] \right\}^2 f(\varepsilon)d\varepsilon;$$
$$(21)$$
$$dw_a = \frac{\pi}{2} \frac{\delta(\varepsilon_{\parallel e}^{'} - \omega_e - \varepsilon_{\parallel} + s\Omega)}{\varepsilon\varepsilon_{e,a}^{'}} \sum_{n=0} |a_n|^2 \frac{n!}{(n+s)!} e^{-\alpha} \left\{ e(\mathbf{A}_{02}\mathbf{p}_{\parallel})\alpha^{3/2}L_{n+s}^3(\alpha) \right.$$
$$\left. + \frac{1}{\sqrt{2}} \frac{e(A_{02})_z}{z_0} [(n+1)\alpha^{(s-1)/2}L_{n+s}^{s-1}(\alpha) + \alpha^{(s+1)/2}L_{n+s}^{s+1}(\alpha)] \right\}^2 f(\varepsilon)d\varepsilon$$

(the superscript s of the probabilities indicates that the processes considered are accompanied by transitions through $s = |n-m|$ levels in the oscillator potential for the transverse motion of the particle motion; $\alpha \equiv (k_{\perp}z_0)^2/2$, where $(n\alpha)^{1/2} \sim (\varepsilon_{\perp}/\varepsilon)^{1/2}s\gamma^2\theta < 1$ is the characteristic oscillator length

In the case of large quantum numbers n, we use the Laguerre polynomials [9]:

$$L_n^s(\alpha) \approx \frac{\Gamma(n+s+1)}{n!} e^{\alpha/2} (n\alpha)^{-s/2} J_s(2\sqrt{n\alpha}) \quad (n \gg 1), \tag{22}$$

where $J_s(2\sqrt{n\alpha})$ is a Bessel function.

We shall distinguish hereafter between two limiting cases, when $(n\alpha)^{1/2} < 1$ and $(n\alpha)^{1/2} \gg 1$. The inequality $(n\alpha)^{1/2} \sim (\varepsilon_\perp/\varepsilon)^{1/2} s\gamma^2\theta < 1$ corresponds to the dipole approximation and is realized in a collinear geometry, when the traveling wave propagates in the direction of the longitudinal of the particle. In the dipole approximation, the asymptotic behavior of the function (22) is given by [9]

$$L_n^s(\alpha) \approx n^s e^{\alpha/2} / s! \quad ((n\alpha)^{1/2} < n, \quad n \gg s),$$

with the aid of which we obtain from (21)

$$\begin{aligned}
dw_e &= \frac{\pi}{2} e(\mathbf{A}_{02})^2 \frac{\delta(\varepsilon'_{\|e} \pm \omega_e - \varepsilon_\| - s\Omega)}{\varepsilon\varepsilon'_e} \sum_{n=s} |a_n|^2 n \frac{(n\alpha)^{s-1}}{[(s-1)!]^2} \\
&\times \left\{ |\mathbf{e}_2 \mathbf{e}_p| \varepsilon \frac{\alpha^{1/2}}{s} + \frac{|\mathbf{e}_2 \mathbf{e}_z|}{\sqrt{2}z_0}\left[1 + \frac{n\alpha}{s(s+1)}\right] \right\}^2 f(\varepsilon) d\varepsilon; \\
dw_a &= \frac{\pi}{2} e(\mathbf{A}_{02})^2 \frac{\delta(\varepsilon'_{\|e} - \omega_e - \varepsilon_\| + s\Omega)}{\varepsilon\varepsilon'_a} \sum_{n=0} |a_n|^2 (n+s) \frac{(n\alpha)^{s-1}}{[(s-1)!]^2} \\
&\times \left\{ |\mathbf{e}_2 \mathbf{e}_p| \varepsilon \frac{\alpha^{1/2}}{s} + \frac{|\mathbf{e}_2 \mathbf{e}_z|}{\sqrt{2}z_0}\left[1 + \frac{n\alpha}{s(s+1)}\right] \right\}^2 f(\varepsilon) d\varepsilon,
\end{aligned} \tag{23}$$

where $\mathbf{e}_p = \mathbf{p}_\| / p_\|$; $\mathbf{e}_z$ is the unit vector of the Cartesian axis Z. From (23) it follows, particular, that for a wave with vector &, perpendicular to the Z axis the probabilities of emission and absorption of a quantum vanish in the limit as $\alpha \to 0$.

In the case of arbitrary polarization of the wave, the first term in the expressions in curly brackets in (23) is small compared with the second to the extent that $((n\alpha)^{1/2} s)(\varepsilon_\perp / \varepsilon)^{1/2}\theta \sim \gamma^2\theta^2 \ll 1$. Thus (we confine ourselves hereafter to the case of the lowest value s = 1).

$$dw_e^{(s=1)} = \frac{\pi}{2}\left(\frac{eA_{02}}{z_0}\right)^2 |\mathbf{e}_2\mathbf{e}_z|^2 \frac{\delta(\varepsilon'_{\|e} + \omega_1 - \varepsilon_\| - \Omega)}{\varepsilon\varepsilon'_e} \sum_{n=0} |a_n|^2 \, nf(\varepsilon)d\varepsilon,$$

$$dw_a^{(s=1)} = \frac{\pi}{4}\left(\frac{eA_{02}}{z_0}\right)^2 |\mathbf{e}_2\mathbf{e}_z|^2 \frac{\delta(\varepsilon'_{\|e} - \omega_1 - \varepsilon_\| + \Omega)}{\varepsilon\varepsilon'_a} \sum_{n=0} |a_n|^2 \, (n+1)f(\varepsilon)d\varepsilon. \quad (24)$$

The averaging over $\varepsilon$ in Eqs. (23) and (24) is easily carried out using the conditions $\Omega \ll \omega \ll \varepsilon_\| \approx \varepsilon \approx |\mathbf{p}|$. In this approximation the $\delta$ functions can be represented in the form

$$\delta(\varepsilon'_{\|e} \pm \omega_s - \varepsilon_\| \mp s\Omega) = \frac{\delta(\varepsilon - \varepsilon_{e,a})}{|\partial\varepsilon'_{\|e,a}/\partial\varepsilon_\| - 1|} = \frac{\gamma^2\varepsilon\delta(\varepsilon - \varepsilon_{e,a})}{\omega_s}, \quad (25)$$

where $\varepsilon_{e,a} = \varepsilon_0 \pm \Delta\varepsilon$ are the energies of the electrons that emit or absorb a quantum of energy $\omega_s$ at a given oscillation frequency $\Omega$;

$$\varepsilon_0 = m_e(\omega_s / 2s\Omega)^{1/2}, \qquad \Delta\varepsilon = \omega_s / 2. \quad (26)$$

## 4. CONCLUSION

We turn to an analysis of basic expressions obtained in the paper. As follows from (19), amplification of the wave at the frequency $\omega_1 = 2\gamma^2\Omega$ is possible in the collinear scheme. When a wave is launched at an angle $\theta$ to the direction $\mathbf{p}_\|$ if $(\varepsilon_\perp/\varepsilon)^{1/2}\gamma^2\theta \geq 1$ amplification is possible at the frequency $\omega_1 = 2\gamma^2 s\Omega$ $(\gamma^2\theta^2 \ll 1)$.

We investigate induced emission of ultrarelativistic electrons in strong electric (magnetic) fields that are uniform along the direction of the electron motion and are not uniform in the transverse direction.

The calculation in the present paper was performed for a harmonic dependence of the potential on the transverse coordinate of the particle. A separate analysis is necessary to cast light on the role of anharmonicity.